\documentclass[referee,useAMS,usenatbib]{mn2e}
\usepackage{graphicx,graphics,amsmath,amssymb,mathrsfs}
\newcommand{\nin}{\noindent}
\newcommand{\dmath}{\mbox{${\mathrm d}$}}
\newcommand{\ddr}{\frac{\dmath}{\dmath R}}

\newcommand{\der}[2]{\frac{\dmath{#1}}{\dmath{#2}}}
\newcommand{\DER}[2]{\frac{\partial{#1}}{\partial{#2}}}
\newcommand{\dder}[2]{\frac{\dmath^2{#1}}{\dmath{#2}^2}}

\newcommand{\pom}{\dot{\varpi}}
\newcommand{\per}[2]{{#1}_{a}^{1{#2}}}
\newcommand{\sigd}[1]{\Sigma_{d}^{#1}} 
\newcommand{\siga}[2]{\Sigma_{#1}^{#2}}
\newcommand{\bfpar}[3]{{\bf{#1}}_{#2}^{#3}}

\begin{document}

\title{Unstable $m=1$ modes of counter--rotating Keplerian discs}

\author[Gulati, Saini \& Sridhar]{Mamta Gulati$^{1,2,3}$, Tarun Deep Saini$^{1,   
  4}$ and S. Sridhar$^{2,5}$\\
  $^{1}$ Indian Institute of Science, Bangalore 560 012, India\\
  $^{2}$ Raman Research Institute, Sadashivanagar, Bangalore 560 080, India\\
  $^{3}$ mgulati@rri.res.in\,,
  $^{4}$ tarun@physics.iisc.ernet.in\,,	
  $^{5}$ ssridhar@rri.res.in 
}
\maketitle

\begin{abstract}
We study the linear $m=1$ counter-rotating instability in a
two--component, nearly Keplerian disc. Our goal is to understand these
\emph{slow} modes in discs orbiting massive black holes in galactic
nuclei. They are of interest not only because they are of large
spatial scale---and can hence dominate observations---but also because
they can be growing modes that are readily excited by accretion
events. Self--gravity being nonlocal, the eigenvalue problem results
in a pair of coupled integral equations, which we derive for a
two-component softened gravity disc. We solve this integral eigenvalue
problem numerically for various values of mass fraction in the
counter-rotating component. The eigenvalues are in general complex,
being real only in the absence of the counter--rotating component, or
imaginary when both components have identical surface density
profiles. Our main results are as follows: (i) the pattern speed
appears to be non negative, with the growth (or damping) rate being
larger for larger values of the pattern speed; (ii) for a given value
of the pattern speed, the growth (or damping) rate increases as the
mass in the counter--rotating component increases; (iii) the number of
nodes of the eigenfunctions decreases with increasing pattern speed
and growth rate. Observations of lopsided brightness distributions
would then be dominated by modes with the least number of nodes, which
also possess the largest pattern speeds and growth rates.
\end{abstract}

\begin{keywords}
instabilities --- stellar dynamics --- celestial mechanics --- galaxies: nuclei
\end{keywords}

\section{Introduction}

Most galactic nuclei are thought to host massive black holes and dense
clusters of stars whose structure and kinematics are correlated with
global galaxy properties \citep{geb96, fm00,geb00}. Such correlations
raise questions of great interest related to galaxy formation and the
growth of nuclear black holes. The nearby large spiral galaxy M31 has
an off--centered peak in a double--peaked brightness distribution
around its nuclear black hole \citep{lds74, lau93, lau98, kb99}. This
lopsided brightness distribution could arise naturally if the apoapses
of many stellar orbits, orbiting the black hole, happened to be
clustered together \citep{tre95}. Since then, kinematic and dynamical
models of such an eccentric disc have been constructed by several
authors \citep{bac01, ss01, ss02, pt03}. Of particular interest to
this work is the model of \citet{ss02}, which included a few per cent
of stars on retrograde (i.e. counter--rotating) orbits. They proposed
that these stars could have been accreted to the centre of M31 in the
form of a globular cluster that spiraled in due to dynamical
friction. This proposal was motivated by the work of \citet{tou02},
who demonstrated that a Keplerian axisymmetric disc is susceptible to
a linear lopsided instability in the $m = 1$ mode, even when a small
fraction of the disc mass is in retrograde motion.

\citet{tou02} considered the linearized secular dynamics of particles
orbiting a point mass, wherein particle orbits may be thought of as
slowly deforming elliptical rings of small eccentricities. The $m=1$
counter--rotating instability was studied analytically for a two--ring
system, and numerically for a many--ring system. The corresponding
problem for continuous discs was then studied by \citet{st10}, who
proposed a simple model with dynamics that could be studied largely
analytically in the Wentzel--Kramers--Brillouin (WKB)
approximation. Their model consisted of a two--component
\emph{softened} gravity disc, orbiting a massive central black
hole. Softened gravity was introduced by \citet{mil71} to simplify the
analysis of the dynamics of stellar systems. In this form of
interaction, the Newtonian $1/d$ gravitational potential is replaced
by $1/\sqrt{d^2 + b^2}$, where $b>0$ is called the {\it softening
  length}. In the context of waves in discs, it is well known that the
softening length mimics the epicyclic radius of stars on nearly
circular orbits \citep{bt08}. Therefore, a disc composed of cold
collisionless matter, interacting via softened gravity, provides a
surrogate for a \emph{hot} collisionless disc.

\citet{st10} used a short--wavelength (WKB) approximation, derived
analytical expressions for the dispersion relation and showed that the
frequency $\omega$ is smaller than the Keplerian orbital frequency by
a factor proportional to the small quantity $\varepsilon = M_d/M\,$
(which is the ratio of the disc mass to mass of the central object);
in other words, the modes are \emph{slow}.  The WKB dispersion
relation was used to argue that equal mass counter--rotating discs
with the same surface density profiles (i.e. when there is not net
rotation) could have unstable modes. They also argued that, for an
arbitrary mass ratio, the discs must be unrealistically hot to avoid
an instability. \citet{st10} then used Bohr--Sommerfeld quantization
to construct global modes, within the WKB approximation. A matter of
concern is that the wavelengths of the modes could be of order the
scale length of the discs; the modes being large--scale it is possible
that the WKB approximation itself is invalid. Another limitation is
that \citet{st10} could construct (WKB) global modes only for the case
of equal mass discs. Therefore it is necessary to address the full
eigenvalue problem to understand the systematic behaviour of
eigenvalues and eigenfunctions. To this end, we formulate the
eigenvalue problem for the linear, slow, $m=1$ modes in a
two--component, softened gravity, counter--rotating disc. Due to the
long--range nature of gravitational interactions, we have to deal with
a pair of coupled integral equations defining the eigenvalue
problem. We draw some general conclusions and then proceed to solve
the equations numerically for eigenvalues and eigenfunctions.

In \S~2 we introduce the unperturbed two--component nearly Keplerian
disc, define the apse precession rates, discuss the potential theory
for softened gravity, and derive the coupled, linear integral
equations that determine the eigenvalue problems for slow $m=1$
modes. A derivation of the relationship between the softened Laplace
coefficients (used in the potential theory of \S~2 and \S~3) and the
usual (unsoftened) Laplace coefficients is given in the Appendix. We
specialize to discs with similar surface density profiles in \S~3,
when the two coupled equations can be cast as a single integral
equation in a new mode variable; this allows us to draw some general
conclusions about the eigenvalue problem. We also discuss in detail
the numerical method to be employed. Our results are presented in
\S~4, where the properties of the stable, unstable and overstable
modes are discussed. Conclusions are offered in \S~5, where we seek to
provide a global perspective on the correlations that occur between
the pattern speeds, growth rates and eigenfunctions, as well as the
variations of these quantities on the mass fraction in retrograde
orbits.
 
\section{Formulation of the linear eigenvalue problem}
We consider linear non--axisymmetric perturbations in two
counter--rotating discs orbiting a central point mass $M$. The discs
are assumed to be coplanar and consist of cold collisionless particles
which attract each other through softened gravity. However, the
central mass and the disc particles interact via the usual
(unsoftened) Newtonian gravity. Softened gravity is known to mimic the
effects of velocity dispersion, so our discs are surrogates for hot
stellar discs. We assume that the total mass in the discs, $M_d$, is
small in comparison to the central mass. Since $\varepsilon \equiv
M_d/M \ll 1$, the dynamics is dominated by the Keplerian attraction of
the central mass, and the self--gravity of the discs is a small
perturbation which enables slow modes. Below we formulate the linear
eigenvalue problem of slow modes.

\subsection{Unperturbed discs}
We use polar-coordinates $\mathbf r\equiv (R\,,\phi)$ in the plane of
the discs, with the origin at the location of the central
mass. Throughout this paper the superscripts `$+$' and `$-$' refer to
the prograde and retrograde components, respectively. The unperturbed
discs are assumed to be axisymmetric with surface densities
$\sigd{\pm}(R)$. The disc particles orbit in circles with velocities,
${\mathbf v}_{d}^{\pm} = \pm\,R\Omega(R) {\mathbf e}_{\phi}$, where
$\Omega(R) > 0$ is the angular speed determined by the unperturbed
gravitational potential,

\begin{equation}
\Phi(R) \;=\; -\,\frac{GM}{R} \;+\; \Phi_{d}(R)\,.
\end{equation}

\nin 
The first term on the right side is the Keplerian potential due
to the central mass, and $\Phi_d(R)$ is the softened gravitational
potential due to the combined self-gravities of both the discs:

\begin{equation}
\Phi_{d}(\mathbf r) \;=\; -\,G\int\,\frac{\sigd{+}({\mathbf
r}') + \sigd{-}({\mathbf r}')}{\sqrt{\left|{\mathbf r}\,-\,{\mathbf r}'\right|^2+b^2}}\;{\mathrm{d}} ^2r' \,,
\label{eq:miller}
\end{equation}

\nin
 where $b$ is the Miller softening length; the potential
$\Phi_d(R)$ is $O(\epsilon)$ compared to $GM/R$. Test particles for
nearly circular prograde orbits have azimuthal and radial frequencies,
$\Omega$ and $\kappa$, given by

\begin{align}
\Omega^{2}(R) &\;=\; \frac{GM}{R^3} \;+\; \frac{1}{R}\der{\Phi_d}{R},\\[1em]
\kappa^{2}(R) &\;=\; \frac{GM}{R^3} \;+\; \frac{3}{R}\der{\Phi_d}{R} \;+\; 
\dder{\Phi_d}{R}\,.
\end{align}

\nin
 The line of apsides of a nearly circular eccentric particle orbit
of angular frequency $\pm\Omega(R)$, subjected only to gravity,
precesses at a rate given by $\pm\pom(R)$, where

\begin{align}
\pom(R) &\;=\; \Omega(R) \;-\; \kappa(R)  \nonumber\\[1em]
&\;=\; -\frac{1}{2\Omega(R)}\left(\frac{2}{R}\ddr + \dder{}{R}\right)\Phi_d(R)
\;+\; O(\varepsilon^2)\,.
\label{eq:pom}
\end{align}

\nin 
The cancellation of the $O(1)$ term, $(GM/R^3)$, which is common
to both $\Omega^2$ and $\kappa^2$ makes $\pom\sim
O(\varepsilon)$. This is the special feature of nearly Keplerian discs
which is responsible for the existence of (slow) modes whose
eigenfrequencies are $\sim O(\varepsilon)$ when compared with orbital
frequencies.

\subsection{Perturbed discs}
Let $\bfpar{v}{a}{\pm}({\mathbf r}, t) = u_{a}^{\pm}(\mathbf{r}{}{},t)
{\mathbf e_R} + v_{a}^{\pm}(\mathbf{r}{}{},t) {\mathbf e_{\phi}}$ and
$\siga{a}{\pm}({\mathbf{r}},t)$ be infinitesimal perturbations to the
velocity fields and surface densities of the $\pm$ discs,
respectively.  These satisfy the following linearized Euler and
continuity equations:

\begin{align}
\DER{\bfpar{v}{a}{\pm}}{t} \;+\; & (\bfpar{v}{d}{\pm}\cdot\nabla)\bfpar{v}{a}{\pm} \;+\; (\bfpar{v}{a}{\pm}\cdot \nabla)\bfpar{v}{d}{\pm} \;=\; -\nabla\Phi_a\,,\label{eq:per1}\\[1em]
\DER{\siga{a}{\pm}}{t} \;+\; & \nabla \cdot (\sigd{\pm}\bfpar{v}{a}{\pm} \;+\; \siga{a}{\pm}\bfpar{v}{d}{\pm}) \;=\; 0,
\label{eq:per2}
\end{align}

\nin
 where $\Phi_a({\bf{r}},t)$ is the perturbing potential. Fourier
 analyzing the perturbations in $t$ and $\phi$, we seek solutions of
 the form, $X_{a}({\mathbf r}, t) = \sum_m X_{a}^{m}(R)\exp[{\rm
     i}(m\phi - \omega t)]\,$. Then

\begin{align}
u_{a}^{m\pm} \;=\;& -\frac{\rm i}{D_{m}^{\pm}}\left[(\pm m\Omega - \omega)\ddr
\;\pm\; 
\frac{2m\Omega}{R}\right]\Phi_{a}^{m}\,,\label{eq:per3}\\[1em]
v_{a}^{m\pm} \;=\;& \frac{1}{D_{m}^{\pm}}\left[\pm\frac{\kappa^2}{2\Omega}\ddr \;+\; \frac{m}{R}(\pm m\Omega - \omega)\right]\Phi_{a}^{m}\,,\label{eq:per4}
\\[1em]
{\rm i}(\pm m\Omega -& \omega)\siga{a}{m\pm} \;+\;
\frac{1}{R}\ddr(R\sigd{\pm}u_{a}^{m\pm})
\;+\; \frac{{\rm i}m}{R}\sigd{\pm}v_{a}^{m\pm} = 0\,,\label{eq:per5}
\end{align}

\nin
where

\begin{align}
D_{m}^{\pm} \;=\;& \kappa^2 \;-\; (\pm m\Omega - \omega)^2\,.\label{eq:per6}
\end{align}

\nin
The above equations determine $u_{a}^{m\pm}$, $v_{a}^{m\pm}$ and
$\siga{a}{m\pm}$ in terms of the perturbing potential
$\Phi_{a}^{m}\,$; this would be the solution were $\Phi_{a}^{m}$ due
to an external source.

We are interested in modes for which $\Phi_{a}^{m}$ arises from self
gravity.  In this case it depends on the total perturbed surface
density, $\left[\siga{a}
  {m+}(R)+\siga{a}{m-}(R)\right]\,$. Manipulating the Poisson integral
given in Eq.~\ref{eq:miller}, we obtain

\begin{align}
\Phi_{a}^{m}(R) \;=\;& \int_{0}^{\infty}R'{\mathrm{dR'}}\,P_{m}(R,R')\,[\Sigma_{a}^{m+}(R')+\Sigma_{a}^{m-}(R')]\,,
\label{eq:psn}
\end{align}

\nin
where the kernel

\begin{align}
P_{m}(R,R') \;=\;& -\frac{\pi G}{R_{>}}B_{1}^{(m)}(\alpha,\beta) \;+\; 
\frac{\pi GR}{R'^{2}}(\delta_{m,1} + \delta_{m,-1})\,.
\label{eq:pm}
\end{align}

\nin
The second term on the right side is the indirect term arising from
the fact that our coordinate system can be a non--inertial frame,
because its origin is located on the central mass. The first term is
the direct term coming from the perturbed self--gravity. Here $R_{<} =
\min(R,R')\,$, $R_{>} = \max(R,R')\,$, $\alpha = R_{<}/R_{>}$ and
$\beta = b/R_{>}\,$. The functions,

\begin{align}
B_{s}^{(m)}(\alpha,\beta) \;=\;& \frac{2}{\pi}\int_{0}^{\pi}{\mathrm{d\theta}}\,\frac{\cos m\theta}{(1-2\alpha \cos\theta + \alpha^2 + \beta^2)^{s/2}}\,,
\label{eq:BSM}
\end{align}

\nin
are ``softened Laplace coefficients'', introduced in
\cite{tou02}. They can be expressed in terms of the usual (unsoftened)
Laplace coefficients, as shown in Appendix~A. We note that the
unperturbed disc potential $\Phi_d$ can be obtained from the
unperturbed disc density, $\sigd{+}(R) + \sigd{-}(R)\,$, by using
Eq.~(\ref{eq:psn}) with $m=0\,$.

\subsection{Slow $m = 1$ modes}
Modes with azimuthal wavenumber $m = \pm 1$ are slow in the sense that
their eigenfrequencies, $\omega$, are smaller than the orbital
frequency, $\Omega$, by a factor $\sim O(\varepsilon)\,$. Without loss
of generality we may choose $m=1$. In the slow mode approximation
\citep{tre01}, we use the fact that $\Omega\gg\omega$ in
Eqs.~(\ref{eq:per3})---(\ref{eq:per6}), and write

\begin{align}
u_{a}^{1\pm}  \;=\;& \mp\,\frac{\rm i}{D_{1}^{\pm}}\left[\Omega\ddr \;+\;
\frac{2\Omega}{R}\right]\Phi_{a}^{1}\,,\label{eq:per7}\\[1em]
v_{a}^{1\pm}  \;=\;& \pm\,\frac{1}{D_{1}^{\pm}}\left[\frac{\Omega}{2}\ddr
\;+\; \frac{\Omega}{R}\right]\Phi_{a}^{1}\,,\label{eq:per8}\\[1em]
\pm\,{\rm i}\,\Omega \Sigma_{a}^{1\pm} \;+\;& \frac{1}{R}\ddr(R\Sigma_{d}^{\pm}u_{a}^{1\pm}) \;+\;\frac{\rm i}{R}\Sigma_{d}^{\pm}v_{a}^{1\pm} \;=\; 0\,,
\label{eq:per9}
\end{align}

\nin
where
\begin{align}
 D_{1}^{\pm} \;=\;& \pm 2\Omega(\omega \;\mp\; \pom).\label{eq:per10}
\end{align}

\nin
Eqs.~(\ref{eq:per7}) and (\ref{eq:per8}) imply the following relations between the perturbed velocity amplitudes:

\begin{align}
u_{a}^{1\pm} \;=\;& -2 {\rm i}v_{a}^{1\pm}\,,\nonumber\\[1ex]
D_{1}^{-}u_{a}^{1-} \;=\;& - D_{1}^{+}u_{a}^{1+}\,,\nonumber\\[1ex]
D_{1}^{-}v_{a}^{1-} \;=\;& - D_{1}^{+}v_{a}^{1+}\,.
\label{eq:per11}
\end{align}

\nin 
We use Eqs.~(\ref{eq:per11}) in the continuity equation
(\ref{eq:per9}) to eliminate $u_{a}^{1\pm}$ and write,

\begin{equation}
\pm\,\Omega\Sigma_{a}^{1\pm} \;=\; \frac{2}{R^{1/2}}\ddr(R^{1/2}
\sigd{\pm}\per{v}{\pm})\,.
\label{eq:per12}
\end{equation}

\nin
Combining Eqs.~(\ref{eq:psn}), (\ref{eq:per8}) and (\ref{eq:per10})---(\ref{eq:per12}) we obtain
 
\begin{align}
\left[\omega \;\mp\; \pom(R)\right]\per{v}{\pm}(R) \;=\;&  
\int_{0}^{\infty}\frac{{\mathrm{d}R'}\,R'^{1/2}}{2R^{2}\Omega(R')}\,\left\{\DER{}{R}\left[R^2 P_{1}(R,R')\right]\right\}\times\nonumber\\[1em]
&\quad\left\{\der{}{{R'}} \left[R'^{1/2}\sigd{+}(R')\per{v}{+}(R') \;-\; 
R'^{1/2}\sigd{-}(R')\per{v}{-}(R')\right] \right\}\,.
\label{eq:per13}
\end{align}

\nin
We rewrite this by defining

\begin{equation}
z^{\pm}(R) \;=\; \left[\frac{R^2\sigd{\pm}(R)}{\Omega(R)}\right]^{1/2}\,
\per{v}{\pm}\,,
\label{zpmdef}
\end{equation}

\nin
use the fact that $\Omega(R) \propto R^{-3/2}$ for a Keplerian flow, and integrate by parts to obtain,

\begin{align}
\left[\omega \;\mp\; \pom(R)\right]z^{\pm}(R) \;=\;& -\int_{0}^{\infty}\frac{{\mathrm{d}}R'}{R'}\,
2\mathcal{F}(R,R')\left[\frac{\sigd{+}(R')\sigd{\pm}(R)}{\Omega(R')\Omega(R)}\right]^{1/2}z^{+}(R')\nonumber\\[1em]
& + \,\, \int_{0}^{\infty}\frac{{\mathrm{d}}R'}{R'}\,2\mathcal{F}(R,R')
\left[\frac{\sigd{-}(R')\sigd{\pm}(R)}{\Omega(R')\Omega(R)}\right]^{1/2}z^{-}(R')\,,\label{eq:per15}
\end{align}

\nin
where 

\begin{equation}
\mathcal{F}(R,R') \;=\; \left(1+\frac{1}{2}\DER{}{\,\ln R'}\right)\left(1+\frac{1}{2}\DER{}{\,\ln R}\right)P_{1}(R,R')\,.
\label{eq:F}
\end{equation}

\nin 
It is convenient to write $\sigd{-}(R) = \eta(R)\sigd{}(R)$ and
$\sigd{+}(R)= (1-\eta(R))\sigd{}(R)\,$, where $\eta(R)$ is the local
mass fraction in the unperturbed counter-rotating component; by
definition, $ 0 \le \eta(R) \le 1$.  Then, Eq.~$(\ref{eq:per15})$ can
be recast as

\begin{align}
\nonumber \omega z^{+}(R) \;=\; +\pom z^{+}(R) \;+\;& \int_{0}^{\infty}\frac{{\mathrm{d}}R'}{R'}[(1-\eta(R'))(1-\eta(R))]^{1/2}\mathcal{K}(R,R')z^{+}(R')\\[1em]
\nonumber  -\;& \int_{0}^{\infty}\frac{{\mathrm{d}}R'}{R'}[\eta(R')(1-\eta(R))]^{1/2}\mathcal{K}(R,R')z^{-}(R')\,,\\[2em]
\nonumber \omega z^{-}(R) \;=\; -\pom z^{-}(R) \;+\;& \int_{0}^{\infty}\frac{{\mathrm{d}}R'}{R'}[(1-\eta(R'))\eta(R)]^{1/2}
\mathcal{K}(R,R')z^{+}(R')\\[1em]
-\;& \int_{0}^{\infty}\frac{{\mathrm{d}}R'}{R'}[\eta(R')\eta(R)]^{1/2}\mathcal{K}(R,R')z^{-}(R'),
\label{eq:coupled}
\end{align}

\nin
where the kernel

\begin{align}
\mathcal{K}(R,R') \;=\;& -\,2\left[\frac{\sigd{}(R')\sigd{}(R)}{\Omega(R')\Omega(R)}\right]^{1/2}\,\mathcal{F}(R,R')\nonumber\\[1em]
\;=\;& -\,2\left[\frac{\sigd{}(R')\sigd{}(R)}{\Omega(R')\Omega(R)}\right]^{1/2}\,\left(1+\frac{1}{2}\DER{}{\,\ln R'}\right)\left(1+\frac{1}{2}\DER{}{\,\ln R}\right)P_{1}(R,R')\nonumber\\[1em]
\;=\;& 2\pi G\left[\frac{\sigd{}(R')\sigd{}(R)}{\Omega(R')\Omega(R)}\right]^{1/2}\,\left(1+\frac{1}{2}\DER{}{\,\ln R'}\right)\left(1+\frac{1}{2}\DER{}{\,\ln R}\right)\frac{B_{1}^{(1)}(\alpha,\beta)}{R_{>}}\,.
\label{kerdef}
\end{align}

\nin 
Therefore the kernel $\mathcal{K}(R,R')$ is a real symmetric
function of $R$ and $R'$.\footnote{The contribution from the indirect
  term in $P_1(R,R')$ vanishes, because $\left(2 +
  \partial/\partial\,\ln R'\right)R'^{-2} = 0\,$.}

Using Eqs.~(\ref{eq:per11}) and (\ref{zpmdef}), we can relate the eigenfunctions
 $z^+(R)$ and $z^-(R)$ to each other: 

\begin{equation}
\sqrt{(1-\eta(R))}\left[\omega - \pom(R)\right]z^+(R) \;=\; 
\sqrt{\eta(R)}\left[\omega + \pom(R)\right]z^-(R)\,.
\label{zpmrel}
\end{equation}

\nin 
This relation can, in principle, be used to eliminate one of
$z^+(R)$ or $z^-(R)$ from the coupled Eqs.~(\ref{eq:coupled}), in
which case the eigenvalue problem can be formulated in terms of a
single function (which can be either $z^+(R)$ or $z^-(R)$). However,
such a procedure results in a further complication: the eigenvalue,
$\omega$, will then occur inside the $R'$ integral in the combination,
$(\omega \pm \pom)/(\omega \mp \pom)$, and this makes further analysis
difficult. Eqs.~(\ref{eq:coupled}) are symmetric under the
(simultaneous) transformations, $\left\{`+'\,, \eta(R)\,,
\omega\,\right\} \;\to\; \left\{`-'\,, \left[1-\eta(R)\right]\,,
-\omega\,\right\}$, which interchange the meanings of the terms
prograde and retrograde. It seems difficult to obtain general results
when $\sigd{+}(R)$ and $\sigd{-} (R)$ have different functional
forms. Below we consider the case when the mass fraction, $\eta$, is a
constant; i.e. when both $\sigd{+}(R)$ and $\sigd{-}(R)$ have the same
radial profile.

\section{The eigenvalue problem for constant $\eta$ discs}

When the counter--rotating discs have the same unperturbed surface
density profiles, i.e. $\sigd{+}(R) \propto \sigd{-}(R)\,$, some
general results can be obtained. This case corresponds to the choice
$\eta = \mbox{constant}$, so that $\sigd{-}(R) = \eta\sigd{}(R)$ and
$\sigd{+}(R)= (1-\eta)\sigd{}(R)\,$.  Then the eigenfunctions $z^+(R)$
and $z^-(R)$ are related to each other by,

\begin{equation}
\left[\omega - \pom(R)\right]\sqrt{\eta}\,z^+(R) \;=\; 
\left[\omega + \pom(R)\right]\sqrt{(1-\eta)}\,z^-(R)\,.
\label{zpmnewrel}
\end{equation}

\nin 
Let us define a new function, $Z(R)$, which is a linear combination 
of $z^+(R)$ and $z^-(R)\,$:

\begin{equation}
Z(R) \;=\; \sqrt{1-\eta}\,z^+(R) \;-\; \sqrt{\eta}\,z^-(R)\,.
\label{Zdef}
\end{equation}

\nin
Then equations (\ref{eq:coupled}) can be manipulated to derive a closed 
equation for $Z(R)\,$:

\begin{equation}
\left[\frac{\omega^2 - \pom^2}{(1-2\eta)\omega + \pom}\right]\,Z(R) \;=\;
\int_{0}^{\infty}\frac{\mathrm{d}R'}{R'}\,\mathcal{K}(R,R')\,Z(R')\,, 
\label{Zeqn}
\end{equation}

\nin 
We note that, in this integral eigenvalue problem for the single
unknown function $Z(R)\,$, the (as yet undetermined) eigenvalue
$\omega$ occurs outside the integral. Once the problem is solved and
$Z(R)$ has been determined, we can use Eq.~(\ref{zpmnewrel}) and
(\ref{Zdef}) to recover $z^{\pm}(R)\,$:

\begin{equation}
z^+(R) \;=\; \sqrt{1-\eta}\,\frac{\omega + \pom(R)}{(1-2\eta)\omega + \pom}\,Z(R)\,,\qquad
z^-(R) \;=\; \sqrt{\eta}\,\frac{\omega - \pom(R)}{(1-2\eta)\omega + \pom}\,Z(R)\,.
\label{zpmZexp}
\end{equation}

\nin
Some general conclusions can be drawn:

\begin{enumerate}
\item In Eq.~(\ref{Zeqn}), the kernel $\mathcal{K}(R,R')$ is real
  symmetric.  Therefore, the eigenvalues, $\omega$, are either real or
  come in complex conjugate pairs.

\item When $\eta = 0$, the counter--rotating component is absent,
  which is the case studied by \citet{tre01}; then the left side of
  Eq.~(\ref{Zeqn}) becomes $\left(\omega - \pom\right)Z\,$. Since the
  kernel $\mathcal{K}(R,R')$ is real symmetric, the eigenvalues
  $\omega$ are real, so the slow modes are stable and oscillatory in
  time. Then the eigenfunctions, $Z(R)$ may be taken to be
  real. Therefore $z^+(R) = Z(R)$ is a real function, and $z^-(R) =
  0\,$.
\footnote{When $\eta=1$, the eigenvalues, $\omega$, are again real,
  with $z^-(R) = Z(R)$ a real function, and $z^+(R) = 0\,$.}

\item When $\eta = 1/2\,$, there is equal mass in the
  counter--rotating component, and the surface densities of the $\pm$
  discs are identical to each other. This case may also be thought of
  as one in which there is no net rotation at any radius. Then
  Eq.~(\ref{Zeqn}) becomes

\begin{equation}
\pom^{-1}(R)\left(\omega^2 - \pom^2(R)\right)Z(R) \;=\;
\int_{0}^{\infty}\frac{\mathrm{d}R'}{R'}\,\mathcal{K}(R,R')\,Z(R')\,, 
\label{Zhalfeqn}
\end{equation}

\nin 
Since the kernel $\mathcal{K}(R,R')$ is real symmetric,
$\omega^2$ must be real.  There are two cases to consider, when the
eigenvalues, $\omega$, are either real or purely imaginary.

$\bullet$ When $\omega$ is real, the slow modes are stable and
oscillatory in time.  The eigenfunctions $z^{\pm}(R)$ can be taken to
be real functions.

$\bullet$ When $\omega$ is imaginary, the eigenvalues come in pairs
that are complex conjugates of each other, corresponding to
non--oscillatory growing/damped modes. Let us set $\eta = 1/2$ and
$\omega = \rm i \gamma\,$ (where $\gamma$ is real) in
Eq.~(\ref{zpmZexp}):

\begin{equation}
z^+(R) \;=\; \frac{\rm i \gamma + \pom(R)}{2^{1/2}\pom(R)}\,Z(R)\,,
\qquad\qquad
z^-(R) \;=\; \frac{\rm i \gamma - \pom(R)}{2^{1/2}\pom(R)}\,Z(R)
\end{equation}

\nin 
The function $Z(R)$, which is a solution of Eq.~(\ref{Zhalfeqn}),
can be taken to be a real function multiplied by an arbitrary complex
constant. It is useful to note two special cases: (i) when $Z(R)$ is
purely imaginary, then $z^+(R)$ and $z^-(R)$ are complex conjugates of
each other; (ii) when $Z(R)$ is real, $z^+(R)$ is equal to minus one
times the complex conjugate of $z^-(R)$.
\end{enumerate}

To make progress for other values of $\eta$, it seems necessary to
address the eigenvalue problem numerically; the rest of this paper is
devoted to this.

\subsection{Application to Kuzmin discs}

For numerical explorations of the eigenvalue problem, we consider the
case when both the unperturbed $\pm$ discs are Kuzmin discs, with
similar surface density profiles: $\sigd{-}(R) = \eta\sigd{}(R)$ and
$\sigd{+}(R) = (1-\eta) \sigd{}(R)\,$, where

\begin{equation}
\sigd{}(R) \;=\; \frac{aM_d}{2\pi(R^2 + a^2)^{3/2}}
\end{equation}

\nin 
is the total surface density, $M_d$ is the total disc mass and
$a$ is the disc scale length.  Kuzmin discs, being centrally
concentrated, are reasonable candidates for unperturbed
discs. Moreover, earlier investigations of slow modes
\citep{tre01,st10} have explored modes in Kuzmin discs, so we find
this choice useful for comparisons with earlier work.  The
characteristic values of orbital frequency and surface density are
given by $\Omega^{\star} = \sqrt{{GM}/{a^3}}$, and $\sigd{*} =
{M_d}/{a^2}$, respectively. The coupled Eqs. ~(\ref{eq:coupled}) can
be cast in a dimensionless form in terms of these physical scales. The
net effect is to rescale the eigenvalue $\omega$ to $\sigma$, where

\begin{equation}
\sigma = \left( \frac{\Omega^{\star} a}{G\Sigma^{\star}}\right)   \omega\,.
\label{sigma}
\end{equation}

\nin 
In the following section, all quantities are to be taken as
dimensionless; however, with some abuse of notation, we shall continue
to use the same symbols for them.

\subsection{Numerical method}

Our method is broadly similar to \citet{tre01}. In order to calculate
eigenvalues and eigenfunctions numerically, we approximate the
integrals by a discrete sum using an $N$--point quadrature rule. The
presence of the term ${\mathrm{dR}}/R$ in the integrals suggests that
a natural choice of variables is $u = \log(R)$ and $v = \log(R')$,
where, as mentioned above, $R$ stands for the dimensionless length
$R/a$. The use of a logarithmic scale is numerically more efficient,
because it induces spacing in the coordinate space that increases with
the radius. This handles naturally a certain expected behaviour of the
eigenfunctions: since the surface density in a Kuzmin disc is a
rapidly decreasing function of the radial distance, we expect the
eigenfunctions to also decrease rapidly with increasing
radius. Therefore, discretization of the coupled equations
(\ref{eq:coupled}) follows the schema:
   
\begin{equation}
\int_{0}^{\infty} \frac{\mathrm{d}R'}{R'}\,
\mathcal{K}(R,R')\,z^{+}(R') \quad\longrightarrow\quad \sum_{i=1}^{N}w_{i}\,\mathcal{K}(e^{u},e^{v_i})\,
z^{+}(e^{v_i})\,,
\end{equation}

\nin 
where $w_i$'s are suitably chosen weights. Then the discretized
equations can be written as a matrix eigenvalue problem:

\begin{equation}
\bf{A}\,\zeta \;=\; \sigma\,\zeta\,,\label{eigeqn} 
\end{equation}

\nin
where 

\begin{equation}
\bf{A} \;=\; \left[\begin{array}{cc}
                         (1-\eta)w_j {\mathcal K}_{ij} + \pom_j\delta_{ij} \quad&\quad -\sqrt{\eta(1-\eta)} w_j{\mathcal K}_{ij}\\[1em]
                        \sqrt{\eta(1-\eta)} w_j{\mathcal K}_{ij} \quad&\quad -\eta w_j {\mathcal K}_{ij} - \pom_j\delta_{ij}
                        \end{array}\right]\,,\qquad \mbox{and}\qquad
\bf{\zeta} \;=\; \left(\begin{array}{c}
                         z^{+}_{i}\\[1em] z^{-}_{i}
                        \end{array}\right) \,.
\end{equation}

\nin 
The $2N\,\times\,2N$ matrix $\bf{A}$ has been represented above
in a $2\,\times\,2$ block form, where each of the 4 blocks is a
$N\,\times\,N$ matrix, with row and column indices $i$ and $j$. Note
that $\delta_{ij}$ is the Kronecker delta symbol, and no summation is
implied over the repeated $j$ indices. Thus we have an eigenvalue
problem for eigenvalues $\sigma$, and eigenvectors given by the $2N$
dimensional column vector $\zeta\,$. The use of unequal weights
destroys the natural symmetry of the kernel, but this is readily
restored through a simple transformation given in \S~8.1 of
\citet{prs92}. The grid for our numerical calculations covers the
range $-7 \le \log R \le 5$, which is divided into $N = 4000$ points;
larger values of $N$ give similar results.

We note some differences with \citet{tre01} concerning details of the
numerical method and assumptions. The major difference is in the
treatment of softening: In \citet{tre01}, a dimensionless softening
parameter $\beta = b/R\,$ was introduced, and the eigenvalue problem
for slow modes was solved by holding the parameter $\beta$ constant,
This renders the physical softening length, $b$, effectively dependent
on radius, making it larger at larger radii, thereby not corresponding
to any simple force law between two disc particles located at
different radii. We have preferred to keep $b$ constant, so that the
force law between two disc particles is through the usual Miller
prescription. Other minor differences in treatment are: (i) in
\citet{tre01}, the disc interior to an inner cut-off radius was
assumed to be frozen. In contrast we use a straightforward inner
cut-off radius of $10^{-5}\,$, as mentioned above; (ii) \citet{tre01}
uses a uniform grid in $\log R$, with four--point quadrature in the
intervals between consecutive grid points; we also use a uniform grid
in $\log R$ but instead employ a single $N$--point quadrature for
integration.

Were we dealing with unsoftened gravity (i.e. the case when $b=0\,$),
the diagonal elements of the kernel would be singular. Hence, when the
softening parameter $b$ is much smaller than the grid size, accuracy
is seriously compromised by round-off errors. Typically, the usable
lower limit for $b$ is $\sim 10^{-2}$.

\begin{figure}
\centering
\includegraphics[width=0.8\textwidth]{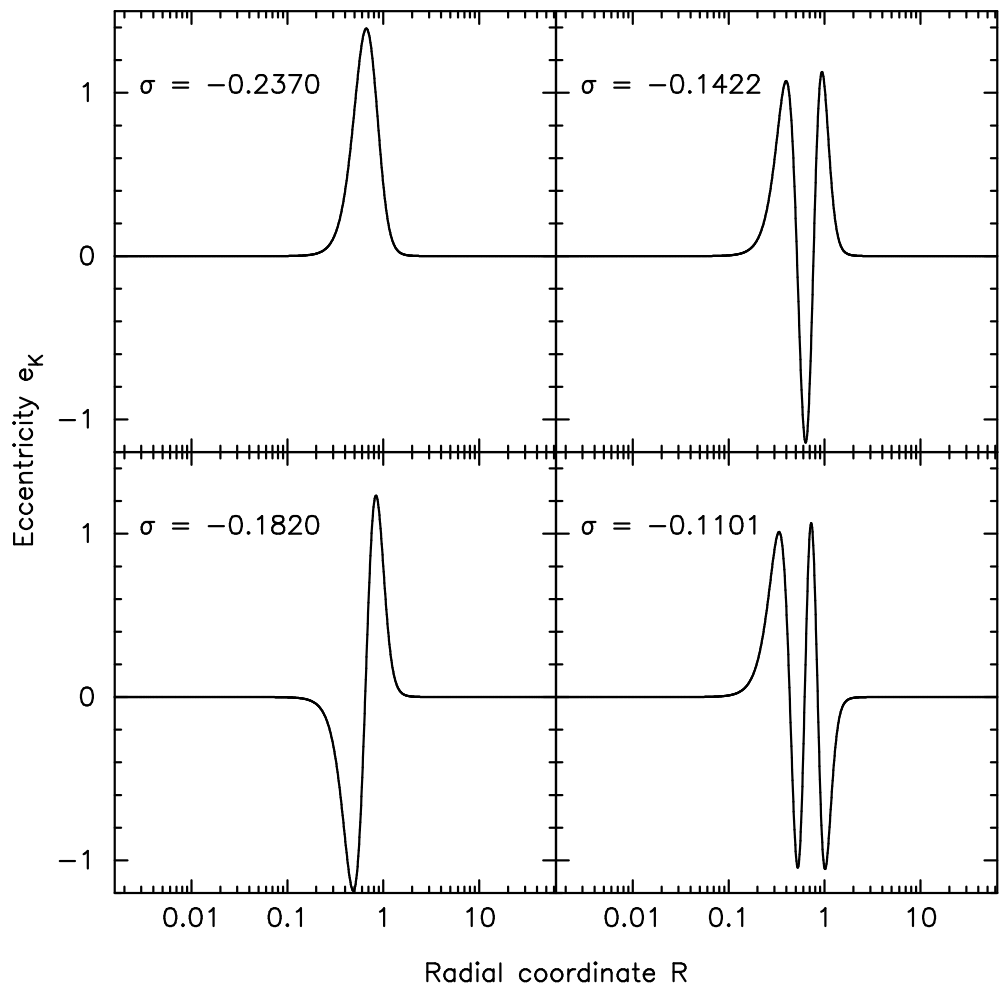}
 \caption{Slow g--modes in a single, prograde ($\eta=0$) Kuzmin disc
   with $\lambda = 0.1$ and $b = 10^{-2}$. The panels are labeled by
   the scaled eigenvalue $\sigma$.}
\label{fig1}
\end{figure}

\begin{figure}
\centering
\includegraphics[width=0.8\textwidth]{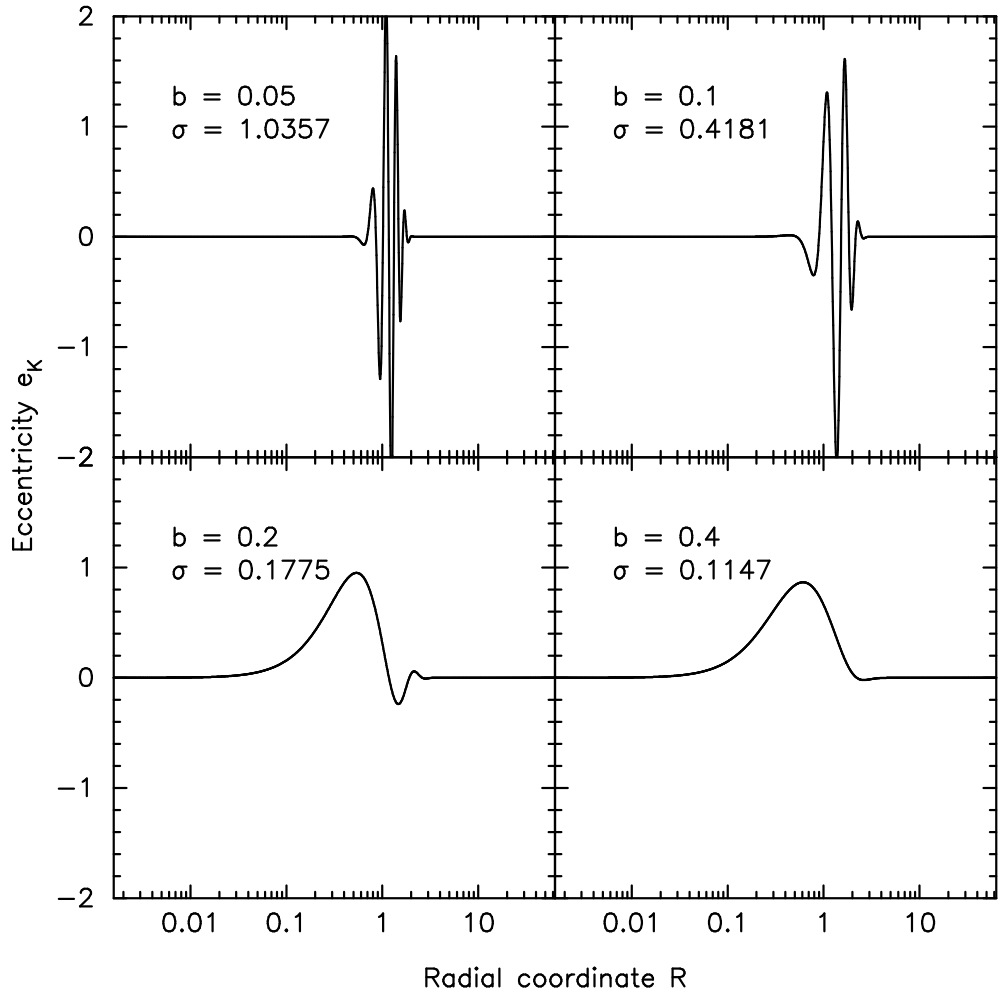}
 \caption{Slow p--modes in single, prograde ($\eta=0$) Kuzmin disc
   with no external source (i.e $\lambda = 1$). The panels are labeled
   by the scaled eigenvalue $\sigma$, and the softening parameter, $b$
   (which has been scaled with respect to $a$, the disc scale
   length).}
\label{fig2}
\end{figure}

\begin{figure}
\centering
\includegraphics[width=0.8\textwidth]{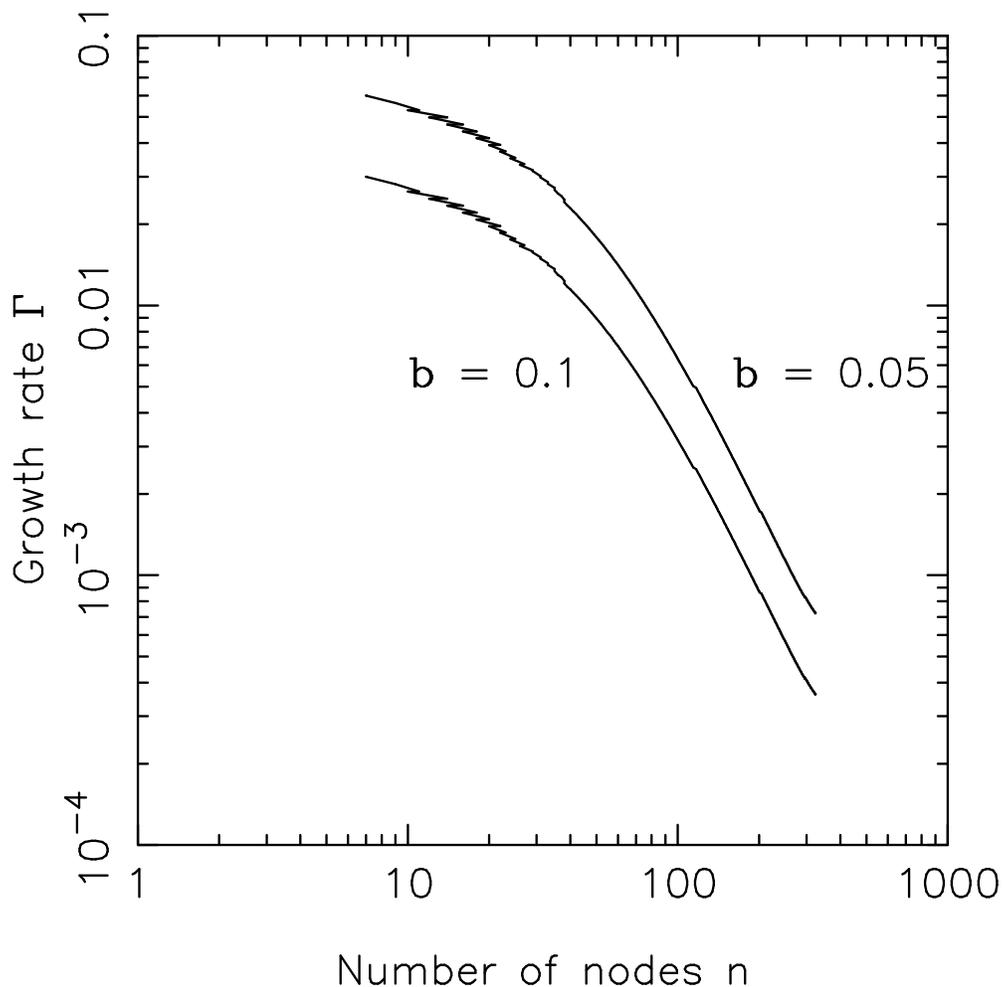}
 \caption{Growth rate versus number of nodes for $\eta = 0.5$, for two
   values of softening, $b=0.1$ and $b=0.05$.}
\label{fig3}
\end{figure}

\begin{figure}
\centering
\includegraphics[width=0.8\textwidth]{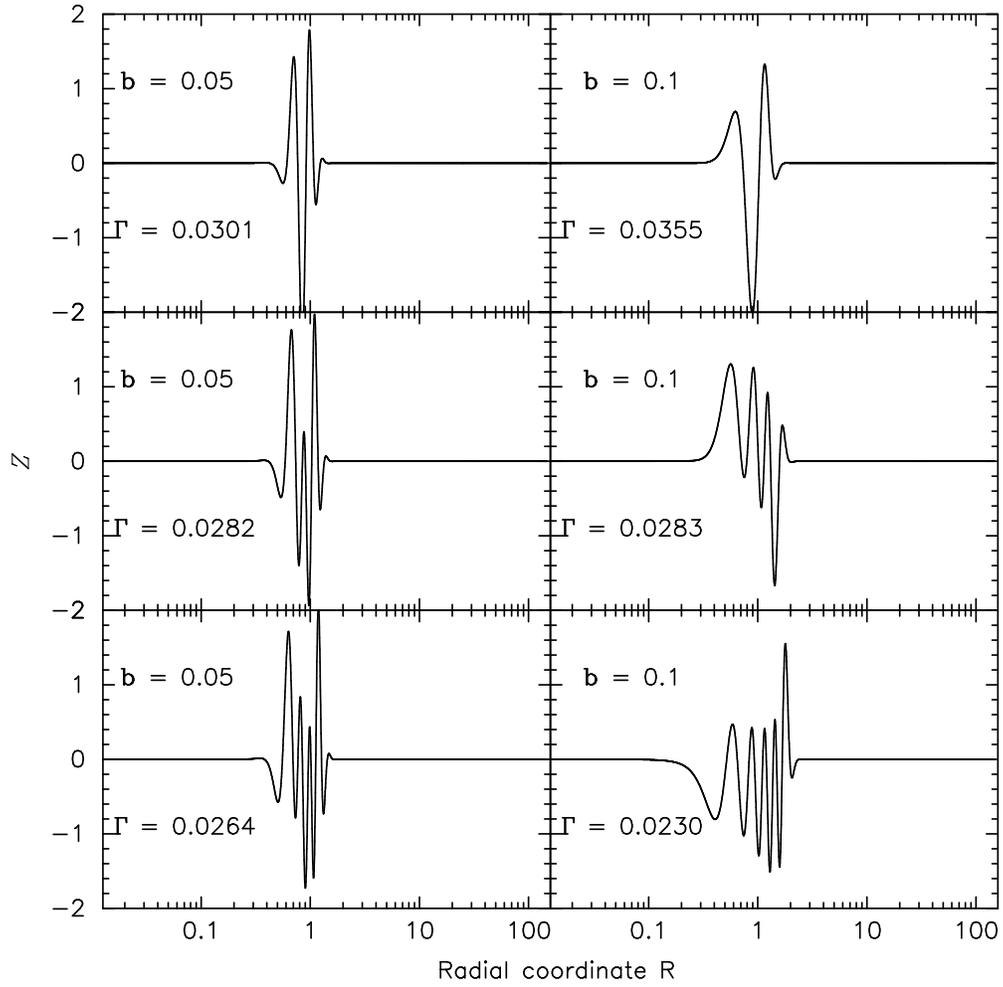}
 \caption{Eigenfunctions $Z(R)$ are plotted as a function of the
   radial coordinate $R$, for $\eta=0.5\,$. The panels are labeled by
   the values of the growth rate, $\Gamma$, and softening, $b$.}
\label{fig4}
\end{figure}

\begin{figure}
\centering
\includegraphics[width=0.8\textwidth]{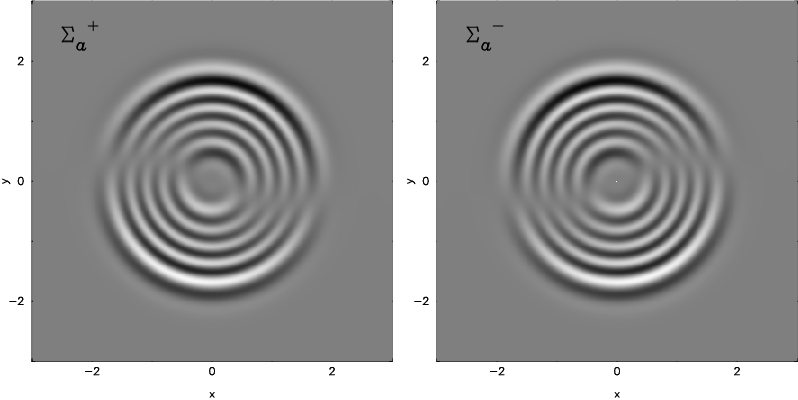}
 \caption{Gray--scale plots of surface density perturbations,
   $\siga{a}{\pm}(R, \phi, t)$ at time $t=0$ for the parameter values,
   $\eta=0.5\,$ and $b=0.1\,$, and $\Gamma = 0.0230$. White/black
   correspond to the maximum positive/negative values of the
   perturbations.}
\label{fig5}
\end{figure}

\section{Numerical results}

We obtained the eigenvalues and eigenfunctions of equation
(\ref{eigeqn}) using the linear algebra package LAPACK
\citep{lapack}. We now present the results of our calculations for
specific values of $\eta$. As noted earlier, interchanging the meaning
of prograde and retrograde orbits leave the results invariant under
the transformation $(\eta,\omega) \to (1-\eta\, ,-\omega)$; therefore,
we present results below only for $0 \le \eta \le 1/2\,$.

\subsection{No counter-rotation: $\eta = 0$}

We are dealing with a single disc whose particles rotate in the
prograde sense.  The eigenvalue problem for this case was studied
first by \citet{tre01}, who also showed that the eigenvalues are real;
in other words, the disc supports stable slow modes. We consider this
case first to benchmark our numerical method as well as assess the
differences in results that may arise due to the manner in which
softening is treated. To facilitate comparison we use the same
nomenclature as \citet{tre01}. Briefly, modes corresponding to
positive and negative eigenvalues are referred to as ``p--modes'' and
``g-modes'', respectively; we also introduce a parameter $\lambda
=(1+f)^{-1}$, where $f$ is a constant that mimics additional
precession due to an external source of the form $\pom_{e}(R) =
f\pom_{d}(R)$; we define eccentricity ${\rm e}_K$ through

\begin{equation}
{\rm e}_K\;=\; 2\,\left(\frac{GM}{R}\right)^{-1/2}\per{v}{}\,,  
\label{eq:ecc}
\end{equation}

\nin
and use the normalization,
 
\begin{equation}
\int \frac{\mathrm{d}\mathrm{R}}{R}\,{\rm e}_K^2(R) \;=\; 1\,.
\end{equation}

Our results corresponding to g--modes, for $\lambda = 0.1$ and $\beta
= 10^{-2}$, are presented in Fig.~\ref{fig1}, where we plot modes with
three or fewer nodes and give their eigenvalues. Results for p--modes
for $\lambda = 1$ (no external source) and various values of softening
parameter $b$, are displayed in Fig.~\ref{fig2}. These figures are to
be compared with Fig.~3 and Fig.~6 of \cite{tre01}: the eigenfunctions
are of broadly similar form, but the eigenvalues differ from those in
\cite{tre01} by upto $\sim 30\%\,$.

\subsection{Equal counter-rotation (or no net rotation): $\eta = 1/2$}

This case was studied by \citet{st10}, who derived the following local
or ``WKB'' dispersion relation:

\begin{equation}
 \omega^2 = \pom\left(\pom + \frac{\pi G\sigd{}(R)}{\Omega(R)}|k|\exp(-|k|b)\right)\,.
 \label{dis:rel}
\end{equation}

\nin 
From this expression they concluded: if $\pom$ happens to be
positive then $\omega$ is real (and the disc is stable), but $\pom$ is
negative for most continuous discs which implies that $\omega$ can be
either real or purely imaginary. \citet{st10} also studied global
modes using Bohr--Sommerfeld quantization, which will be discussed
later in this section.

We have proved in the last section that the eigenvalues are either
real (stable oscillatory modes), or purely imaginary (non--oscillatory
growing and damped modes).  Here we focus on the growing (unstable)
modes; we define the \emph{growth rate} of perturbations as,

\begin{equation}
 \Gamma = \sqrt{\frac{4}{3}} b |\sigma|\,.\label{grthrt}
\end{equation}

\nin 
in order to facilitate comparisons with \citet{st10}. In
Fig.~{\ref{fig3}}, we plot $\Gamma$ for $b$ equal to $0.1$ and $0.05$,
versus the number of nodes of $Z(R)$.  \citet{st10} found two separate
branches in the spectrum, corresponding to long and short wavelengths
(for each value of $b$). Comparing our Fig.~ {\ref{fig3}} with their
Figs.~3 \& 4, we see that our results are more consistent with their
short--wavelength branch than with their long--wavelength branch.
This disagreement is probably because the long--wavelength branch
corresponds to $kR\sim 1$, where WKB approximation breaks
down. Moreover, the agreement between our results and their
short--wavelength branch holds only in a broad sense, because there
are differences in the numerical values of the eigenvalues.  We trace
this difference to the fact that \citet{st10} used an analytical
result for the precession frequency corresponding to unsoftened
gravity, whereas we have consistently used softened gravity for all
gravitational interactions between disc particles. This probably also
results in another difference between our results: according to
\citet{st10}, $10^{-3} < \Gamma < 10^{-2}$; however, as can be seen
from our Fig.~{\ref{fig3}}, we obtain values of $\Gamma$ both inside
and outside this range. Changing the value of $b$ causes a horizontal
shift in the spectrum, which is consistent with their results.

In Fig~{\ref{fig4}}, we plot a few of these eigenfunctions as a
function of $R$ for both values of $b$ equal to $0.1$ and $0.05$. Note
that, from the discussion in the previous section, $Z(R)$ can always
be chosen to be a real function of $R$. The smallest number of nodes
corresponds to the largest value of the growth factor. The
eigenfunctions with the fewest nodes have significant amplitudes in a
small range of radii around $R\sim 1$, and this range increases with
the number of nodes (and correspondingly, the growth rate
decreases). Fig~{\ref{fig5}} is a gray--scale plot of the surface
density perturbations in the $\pm$ discs, $\siga{a}{\pm}(R,\phi,
t=0)$. Note the relative phase shift between the $\pm$
perturbations. For other value of $b$ shown in Fig~{\ref{fig4}} we get
similar patterns.

\begin{figure}
\centering
\includegraphics[width=0.8\textwidth]{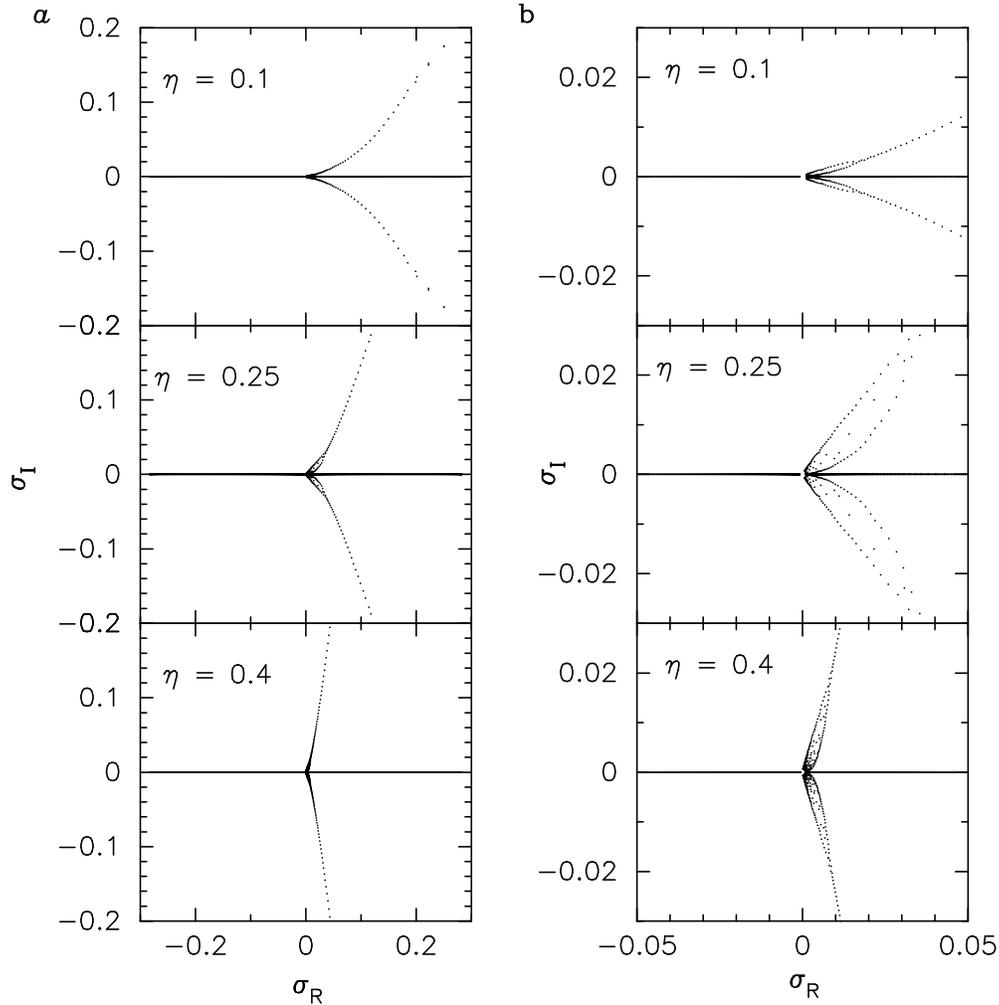}
 \caption{Distribution of eigenvalues in the complex $\sigma$ plane,
   for $\eta = 0.1, 0.25$ and $0.4$. Panels labeled $\mathnormal{a}$
   give an overview, whereas the panels labeled $\mathnormal{b}$
   provide an close--up view near the origin.}
\label{fig6}
\end{figure}

\begin{figure}
\centering
\includegraphics[width=0.8\textwidth]{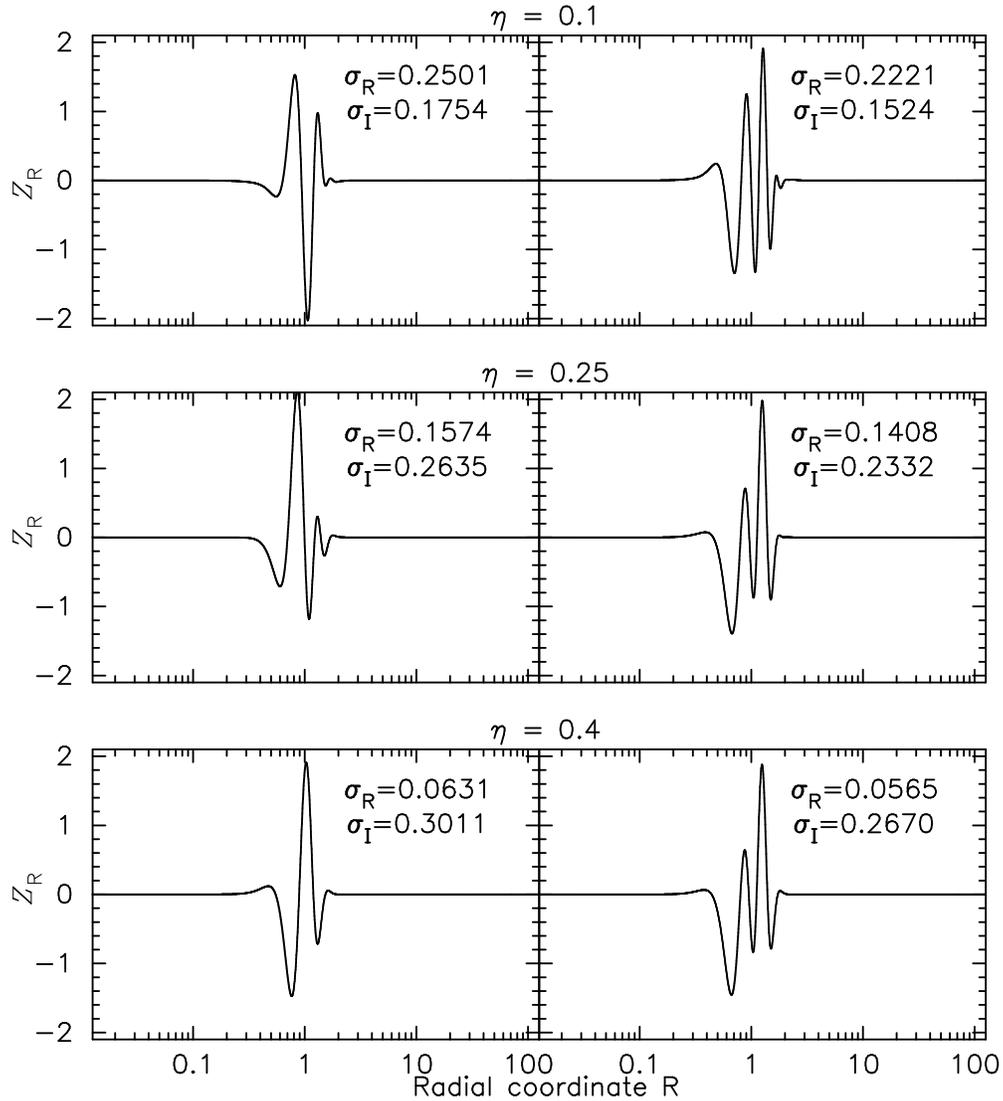}
 \caption{Real parts of the ``most unstable'' eigenfunctions $Z(R)$,
   plotted as a function of the radial coordinate, $R$ for $b=0.1$ and
   for $\eta = 0.1, 0.25$ and $0.4$. Panels are labeled by the real
   and imaginary parts of the eigenvalues.}
\label{fig7}
\end{figure}

\begin{figure}
\centering
\includegraphics[width=0.8\textwidth]{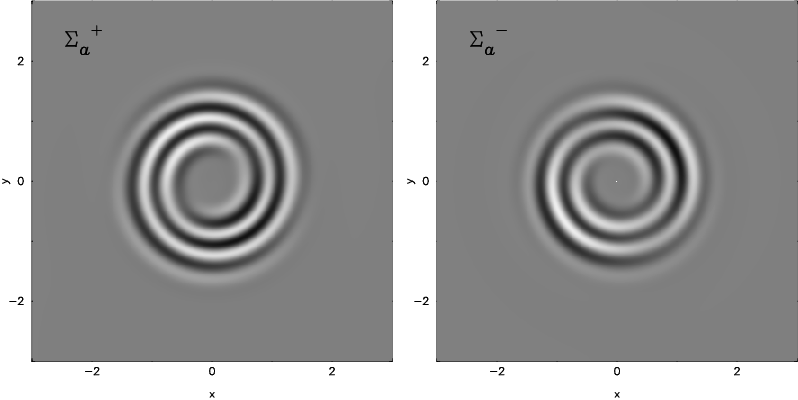}
 \caption{Gray--scale plots of surface density perturbations,
   $\siga{a}{\pm}(R, \phi, t)$ at time $t=0$ for the parameter values,
   $\eta=0.25\,$ and $b=0.1\,$, and $\sigma = 0.1408 + \rm i 0.2332$.
   White/black correspond to the maximum positive/negative values of
   the perturbations.}
\label{fig8}
\end{figure}

\begin{figure}
\centering
\includegraphics[width=0.8\textwidth]{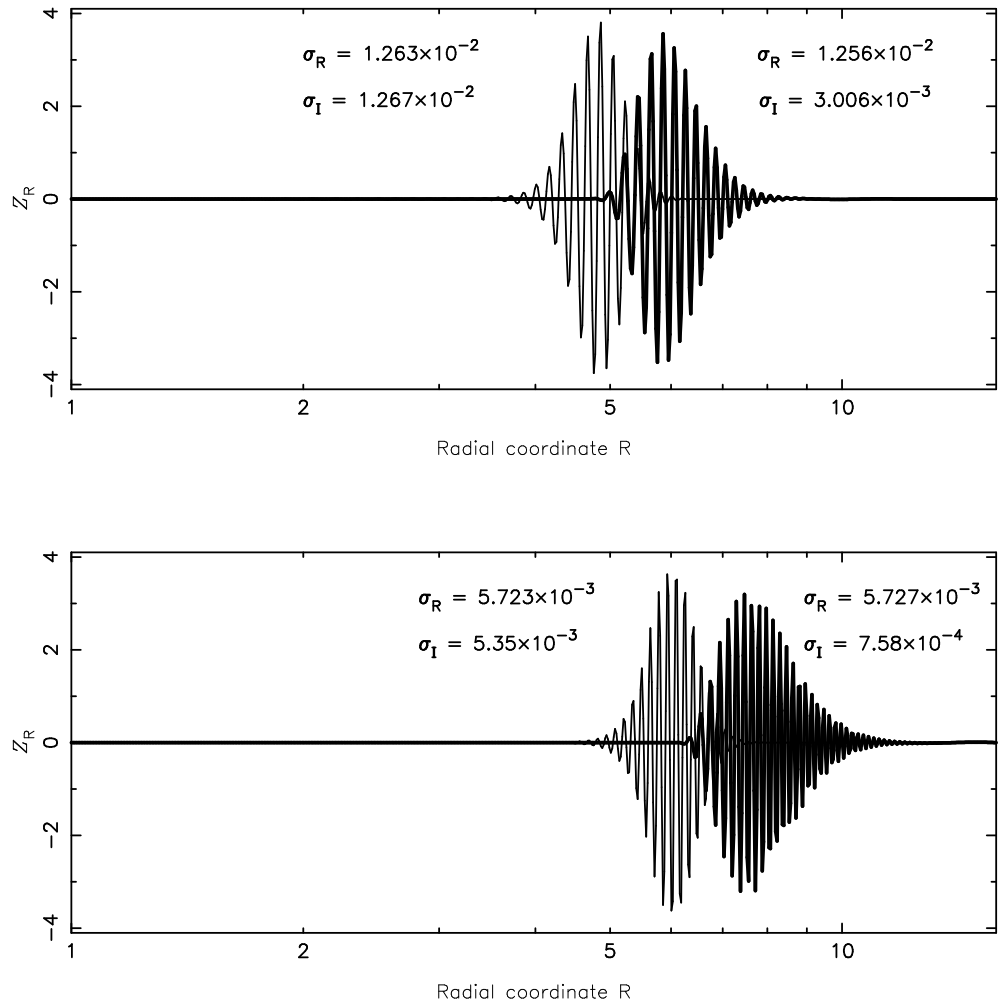}
 \caption{Real parts of two pairs of eigenfunctions $Z(R)$ (from two
   arms of a branch), plotted as a function of the radial coordinate
   $R$, for $b=0.1$ and $\eta = 0.25$. Panels are labeled by the real
   and imaginary parts of the eigenvalues.}
\label{fig9}
\end{figure}

\subsection{Other values of $\eta$}

We present results for values of $\eta$ other than $0$ and
$1/2$. These are particularly interesting, not only because they were
not explored by \citet{st10}, but because the eigenvalues can be truly
complex, corresponding to growing and damped modes which precess with
steady pattern speeds. We write the eigenvalues as $\sigma =
\sigma_{\mathnormal{R}} + \rm i \sigma_{\mathnormal{I}}$. In
Fig~{\ref{fig6}}, we display the eigenvalues in the complex
$\sigma$--plane, for softening parameter $b=0.1$ and for $\eta$ equal
to $0.1$, $0.25$ and $0.4$. Panels on the left, labeled
$(\mathnormal{a})$, provide an overview, whereas the panels on the
right, labeled $(\mathnormal{b})$, provide a close--up view of the
distribution of eigenvalues near the origin of the complex
$\sigma$--plane; this distribution is similar to Fig.~3 of
\citet{tou02}.  We are able to provide much more detail, essentially
because we are dealing with continuous discs rather than a finite
number of rings.

As $\eta$ increases from $0$, the eigenvalues go from real to complex,
a bifurcation that has been traced in \citet{tou02} to a phenomenon
identified by M.~J.~Krein due to the resonant crossing of stable
modes.  The complex eigenvalues come in complex conjugate pairs, so
there are two branches to the distribution. As $\eta$ increases, the
branches progressively separate and, for $\eta = 1/2$ must lie along
the positive and negative imaginary axes. It is intriguing that each
of these two branches consists of more than one arm. In the close--up
views provided by the $(\mathnormal{b})$ panels, it appears as if each
of the branches has two arms; however, more detailed investigations
are required to determine if there are more arms.  The arms of each of
the branches are most widely separated when $\eta=0.25$, which is the
value of $\eta$ exactly midway in its range $0 \leq \eta \leq
0.5\,$. The separations decrease as $\eta$ approaches either $0$ or
$1/2$; this is natural because, for $\eta=0$ both branches must lie on
the real axis and, for $\eta=1/2$ both branches must lie on the
imaginary axis.

The eigenfunctions are in general complex, and have a rich structure
as functions of their eigenvalues. Since our interest is in the
unstable modes, we now display in Fig.~{\ref{fig7}} plots of the $Z_R
= \Re[Z(R)]$ in Eq.~(\ref{Zdef}), corresponding to the ``most
unstable'' modes (for softening parameter $b=0.1$ and for $\eta$ equal
to $0.1$, $0.25$ and $0.4$). In other words, for some chosen value of
$\sigma_R$, we display the real part of the eigenfunction
corresponding to the largest value of $\sigma_{\mathnormal{I}}$. For a
fixed value of $\eta$, the number of nodes of the eigenfunctions
decreases with increasing pattern speed and growth
rate. Fig~{\ref{fig8}} is a gray--scale plot of the surface density
perturbations in the $\pm$ discs, $\siga{a}{\pm}(R,\phi, t=0)$, for
the parameter values $\eta=0.25$ and $b=0.1$, and $\sigma = 0.1408 +
\rm i 0.2332$ .

It is also of interest to ask how eigenfunctions from two different
arms of the same branch behave. To do this, we picked two
eigenfunctions with nearly the same value of $\sigma_R$, but with
values of $\sigma_I$ corresponding to two different arms of one
branch; Fig.~{\ref{fig9}} shows two such pairs of eigenfunctions for
$\eta = 0.25$. We have looked at pairs of such eigenfunctions for
other values of $\eta$, but do not display them, here we note what
seems to be a general trend: (i) the two members of a pair are more
similar to each other when the values of their $\sigma_R$ are closer
to each other; (ii) the member of a pair with the smaller value of
$\sigma_I$ is more displaced toward larger radii.

\section{Conclusions}

We study linear, \emph{slow} $m=1$ modes in softened gravity,
counter--rotating Keplerian discs. The eigenvalue problem is
formulated as a pair of coupled integral equations for the $\pm$
modes. We then specialize to the case when the two discs have similar
surface density profiles but different $\pm$ disc masses. It is of
great interest to study the properties of the modes as a function of
$\eta$, which is the fraction of the total disc mass in the retrograde
population. Recasting the coupled equations as a single equation in a
new modal variable, we are able to demonstrate some general
properties: for instance, when $\eta=1/2$, the eigenvalues must be
purely imaginary or, equivalently, the modes are purely unstable. In
other words, when the $\pm$ discs have identical surface density
profiles then there are growing $m=1$ modes with zero pattern speed, a
conclusion which is consistent with \citet{ara87, pp90, sm94, ljh97,
  tou02, tre05}. To study modes for general values of $\eta$, the
eigenvalue problem needs to be solved numerically. Our method is
broadly based on \citet{tre01}, but there are some differences whose
details have been discussed in the text. The main point of departure
is in the way that softening has been treated. In \citet{tre01}, a
dimensionless softening parameter $\beta = b/R\,$ was introduced, and
the eigenvalue problem was solved by holding the parameter $\beta$
constant. This procedure renders the physical softening length, $b$,
effectively dependent on radius (making it larger at larger radii),
thereby not corresponding to any simple force law between two disc
particles located at different radii. We have preferred to keep $b$
constant, so that the force law between two disc particles is through
the usual Miller prescription.

We calculate eigenvalues and eigenfunctions numerically for discs with
surface density profiles of Kuzmin form. Kuzmin discs, being centrally
concentrated, are reasonable candidates for unperturbed
discs. Moreover, earlier investigations of slow modes
\citep{tre01,st10} have explored modes in Kuzmin discs, so this choice
is particularly useful for comparisons with earlier work. Comparing
our results with those of \citet{tre01} for $\eta=0$ (when the slow
modes are stable), we find that the eigenfunctions are of broadly
similar form, but the eigenvalues differ by up to $\sim 30\%\,$; this
is a result of the different ways in which we have treated
softening. For the case of no net rotation ($\eta=1/2$), we find that
the growth rates (of the unstable modes) we calculate are broadly
consistent with the short--wavelength branch of the global WKB modes
determined earlier by \citet{st10}, but not their long--wavelength
branch. This disagreement probably arises because the long--wavelength
branch corresponds to wavelengths of order the disc scale length where
WKB approximation breaks down. Moreover, the agreement between our
results and their short--wavelength branch holds only in a broad
sense, because there are differences in the numerical values of the
eigenvalues. We trace this difference to the fact that \citet{st10}
used an analytical result for the precession frequency corresponding
to unsoftened gravity, whereas we have consistently used softened
gravity for all gravitational interactions between disc particles.

We have also investigate eigenmodes for values of $\eta$ other than
$0$ and $1/2$. These cases are particularly interesting, not only
because they were not explored by \citet{st10}, but because the
eigenvalues can be truly complex, corresponding to growing (and
damped) modes with non zero pattern speeds. We have presented results
for $\eta = 0.1, 0.25$ and $0.4$ in the previous sections. Based on
these, we interpolate and offer the following conclusions about the
properties of the eigenmodes and their physical implications, for all
values of $\eta$ (which is the mass fraction in the retrograde
population):

\begin{enumerate}  
\item For a general value of $\eta$ (between $0$ and $1/2$), the
  distribution of eigenvalues in the complex plane has two
  branches. These branches are symmetrically placed about the real
  axis, because the eigenvalues come in complex conjugate pairs.

\item The pattern speed appears to be non negative for all values of
  $\eta$, with the growth (or damping) rate being larger for larger
  values of the pattern speed.

\item For a fixed value of $\eta$, the number of nodes of the
  eigenfunctions decreases with increasing pattern speed and growth
  (or damping) rate.

\item For a value of pattern speed in a chosen narrow interval, the
  growth (or damping) rate increases as $\eta$ increases from $0$ to
  $1/2$.

\item Each of the two branches in the complex (eigenvalue) plane has
  at least two arms. When $\eta=0$, the eigenvalues are all real, so
  both branches lie on the real axis, with zero spacing between the
  arms. As $\eta$ increases, the branches lift out of the real axis,
  and the arms separate. It appears as if the maximum separation
  between the arms happens when $\eta=1/4$. As $\eta$ increases
  further, the branches continue to rise with greater slope, while the
  arm separation begins decreasing. Finally, when $\eta=1/2$, the arm
  separation decreases to zero as the branches lie on the imaginary
  axis.
\end{enumerate}

Observations of lopsided brightness distributions around massive black
holes are somewhat more likely to favour the detection of modes with
fewer nodes than modes with a large number of nodes, because the
former suffer less cancellation due to finite angular resolution. From
items (ii) and (iii) above, we note that the modes with a small number
of nodes also happen to be those with larger values of the pattern
speed and growth rate, both qualities that enable detection to a
greater degree. Having said this, it would be appropriate to note some
limitations of our work. Softened gravity discs are, after all,
surrogates for discs composed of collisionless particles (such as
stars) with non zero thickness and velocity dispersions. It is
necessary to formulate the eigenvalue problem for truly collisionless
discs, in order to really deal with stellar discs around massive black
holes. Meanwhile, our results will serve as a benchmark for future
investigations of modes in these more realistic models.

\def\etal{{\it et~al.}}
\def\apj{{Astroph.\@ J. }}
\def\mnras{{Mon.\@ Not.\@ Roy.\@ Ast.\@ Soc.}}
\def\aap{{Astron.\@ Astrophys.}}
\def\aj{{Astron.\@ J.}}
\def\apss{{Astroph.\@ Space \@ Science}}
\def\pasj{{Publ.\@ Astron.\@ Soc.\@ Japan}}
\def\apjl{{Astrophysical.\@ J. {\rm Letters}}}

\appendix
\section{Expressing softened Laplace coefficients in terms of (unsoftened) Laplace coefficients}

Softened Laplace coefficients were defined in \citet{tou02} as,

\begin{equation}
B^m_s(\alpha, \beta) \;=\; \frac{2}{\pi}\int_0^\pi d\theta\,\frac{\cos{m\theta}}{\Delta^{s/2}}
\label{Bdef}
\end{equation}

\noindent
where 

\begin{equation}
\Delta \;=\; 1 + \alpha^2 + \beta^2 -2\alpha\cos\theta
\label{Deltadef}
\end{equation}

\noindent
We now write  

\begin{equation}
\Delta \;=\; \gamma^2 + \delta^2 -2\gamma\delta\cos\theta
\label{Deltaexp}
\end{equation}

\noindent
One solution for $\gamma$ and $\delta$ is,

\begin{eqnarray}
\gamma &\;=\;& \left[\frac{1+\alpha^2+\beta^2}{2} \;+\; \frac{1}{2}\sqrt{(1+\alpha^2+\beta^2)^2 \,-\, 4\alpha^2}\right]^{1/2}\nonumber\\[3ex]
\delta &\;=\;& \frac{\alpha}{\gamma}
\label{gamdeltdef}
\end{eqnarray}

\noindent
Therefore

\begin{equation}
B^m_s(\alpha, \beta) \;=\; \gamma^{-s}\,b^m_{s/2}\left(\delta/\gamma\right)
\label{Bexp}
\end{equation}

\noindent
where

\begin{equation}
b^m_{s/2}(\alpha) \;=\;\frac{2}{\pi}\int_0^\pi d\theta\,\frac{\cos{m\theta}}{\left(1 + \alpha^2 - 2\alpha\cos\theta\right)^{s/2}}
\,;\qquad\mbox{$\alpha < 1$}
\label{bdef}
\end{equation}

\noindent
are the familiar (unsoftened) Laplace coefficients \citep{md99}. From
eqn.~(\ref{Bexp}), we must have $(\delta/\gamma) < 1$.  That this is
indeed true can be proved using eqns.~(\ref{gamdeltdef}): $\gamma$ is
a monotonically increasing function of $\beta^2$, hence $\gamma\geq
1$, and $(\delta/\gamma) = (\alpha/\gamma^2) < 1\,$.
 
\end{document}